\documentstyle[epsf,referee]{laa}
\begin{document}
\thesaurus{08 (09.01.1;  09.04.1;   09.18.1) }           
\title{Upper limit on C$_{60}$ and C$_{60}^+$ features in the ISO-SWS 
spectrum of the reflection
nebula NGC 7023 \thanks{ISO is an ESA project with instruments funded  by ESA 
Member States (especially the PI countries: the United Kingdom, France, 
the Netherlands, 
Germany), and with the participation of ISAS and NASA.}}
\author{C. Moutou (1), K. Sellgren (2), L. Verstraete (3), A. L\'eger (3)}
\offprints{C. Moutou}
\institute{(1) European Southern Observatory\\
Alonso de Cordoba 3107, Santiago, Chile\\
cmoutou@eso.org\\
(2) Astronomy Department, Ohio State University\\
174 West 18th Avenue, Columbus OH 43210, USA \\ 
sellgren@astronomy.ohio--state.edu\\
(3) Institut d'Astrophysique Spatiale, CNRS\\
Universit\'e Paris Sud, b\^atiment 121\\
F-91405 Orsay, France\\
leger@iaslab.ias.fr,verstra@ias.fr\\
}
\date{Received date; accepted date}
\maketitle
\begin{abstract} 
We present here the 7.0 -- 8.7 $\mu$m spectrum of 
the bright reflection nebula NGC 7023. 
Our observations are made with the Short Wavelength Spectrometer (SWS) 
on the European satellite Infrared Space Observatory (ISO). 
The vibrational bands of the ionized fullerene C$_{60}^+$ 
are expected at 7.11 and 7.51 $\mu$m, 
while those of the neutral fullerene C$_{60}$ 
are expected at 7.0 and 8.45 $\mu$m. 
We estimate an upper limit in NGC 7023
for the C$_{60}^+$ abundance of $<$0.26\% of 
the interstellar abundance of carbon, while C$_{60}$ contains
$<$0.27\% of interstellar carbon.
\end{abstract}

\section{C$_{60}$ and related compounds in the interstellar medium}
Kroto et al. (1985) proposed that the fullerene molecule
C$_{60}$ and related species 
play an important role in interstellar physics and chemistry. 
This argument is based on the stability of the ``buckyball''
structure of fullerene compounds, as observed in the laboratory and confirmed 
by ab-initio calculations (see also Kroto 1987, 1988 and references therein). 
The formation mechanisms invoked to produce such spherical 
carbonaceous structures are
usually based on reactions in circumstellar shells, 
with precursors such as acetylene and corannulene  
(Goeres and Seldmayr 1992; Kroto and Jura 1992; Bettens et al. 1997). 
Another way of obtaining fullerenes, together with polycyclic
aromatic hydrocarbons (PAHs), is the photoerosion of larger
hydrogenated amorphous carbon grains (Scott et al. 1997).
Fullerene or fullerane species have been searched for 
at various wavelengths in interstellar spectra. 
They have also been proposed as explanations for many 
unidentified interstellar bands: the extinction bump at 2175 \AA, the
diffuse interstellar bands, the infrared emission features, 
and the extended red emission (Webster 1993a,b,c, 1996).

The C$_{60}$ UV absorption spectrum shows a band
at 3860 \AA\ (Heath et al. 1987).
Snow and Seab (1989) and Somerville and Bellis (1989) have searched for 
this feature without any clear evidence
of its presence. They derived a C$_{60}$ upper limit of 
the order of $<$0.01\% of the cosmic 
carbon abundance in the diffuse medium, 
and approximately $<$0.7\% of the carbon abundance in Mira stars.

More recently, Clayton et al. (1995) have searched for the infrared (IR) 
emission feature of C$_{60}$ at 8.6 $\mu$m 
in the spectra of R Coronae Borealis stars. 
No feature was seen in these carbon-rich, hydrogen-deficient stars 
to a level of 2\% of the continuum. 
According to these authors, only the evolved carbon star IRC+10216 
possibly shows a 8.6 $\mu$m band that may be assigned to C$_{60}$. 

What is the dominant ionization state of C$_{60}$ in the interstellar
environment? We consider recent ionization models for PAHs 
(Bakes and Tielens 1995; Salama et al. 1996; Dartois and d'Hendecourt 1997).
The comparison between PAHs and C$_{60}$
is relevant, as the ionization potentials for C$_{60}$ and PAHs
of the same size are similar (see Table 2 in Leach 1995). 
The models conclude that a 60 atom PAH would be mostly neutral (PAH)
and anionic (PAH$^{-}$) in the diffuse medium, and
cationic (PAH$^{+}$) in more irradiated clouds such as 
reflection nebulae near a hot star. 
Only the model of Bakes and Tielens takes into account the second ionization 
stage (PAH$^{++}$). In the standard PDR model they adopt, 
50\% of the 60 atom species 
would be singly ionized, 35\% would be neutral 
and 15\% doubly ionized. In the case of 
NGC 7023, the radiation field is 10 times fainter 
than in this model, so that the fraction of 
C$_{60}^{++}$ is even lower and we shall ignore it hereafter. 
Laboratory work by Petrie et al. (1993) has shown that
C$_{60}^+$ is the most stable species among 60 atom fullerenes, because of
its ``exceptional unreactivity'' with other interstellar material. 

The cation C$_{60}^+$ has been proposed as a 
diffuse interstellar band (DIB) carrier (L\'eger et al. 1988).
It possesses electronic modes at wavelengths
shorter than 1 $\mu$m, identified by laboratory spectra 
in rare-gas matrices (Fulara et al. 1993), which were expected to be observed 
in absorption in the interstellar medium. 
Foing and Ehrenfreund (1994, 1997) have observed 
absorption features at 9577 and 9632 \AA\,
along a few lines of sight, 
which could correspond to these modes.  
This assignment of the two near-IR
bands to C$_{60}^+$, however, is debated (Maier 1994, Jenniskens et al. 1997) 
and we will discuss this point carefully in section 2. 

The C$_{60}^+$ species would re-emit any energy 
absorbed from a neighboring star through  its four IR modes. 
Two of these modes have been measured in the laboratory to fall
at 7.11 and 7.51 $\mu$m (Fulara et al. 1993). 
The goal of this paper is to search for the vibrational signature 
of C$_{60}^+$ in the high
signal-to-noise spectrum of the reflection nebula NGC 7023. 
We will also search for 
C$_{60}$ bands, measured in the laboratory 
at 7.0 and 8.45 $\mu$m (Kr\"atschmer et al. 1990).

\section{Abundance of C$_{60}^+$ from its electronic transitions}

The laboratory spectrum of C$_{60}^+$ shows two electronic bands at 
9583 and 9645 \AA\ (10435 and 10368 cm$^{-1}$) 
in a Neon matrix (Fulara et al. 1993).
The corresponding absorption bands have been searched for in the 
interstellar medium (ISM),
taking into account the wavelength shift induced by the matrix polarisability. 
The two new interstellar bands  observed at 9577 and 9632 \AA\ (10442 
and 10382 cm$^{-1}$) by Foing and Ehrenfreund (1994, 1997) have been 
tentatively assigned to the cation
fullerene C$_{60}^+$. To deduce the abundance from the 
observations requires
an estimation of the oscillator strength. The most reliable estimates 
($f$ = 0.003 -- 0.006 from Fulara et al. 1993) lead to a total
abundance of 0.6 to 1.2\% of cosmic carbon in this species alone (Moutou et 
al. 1996a), if the C/H ratio is taken as the solar value (3.7 $\times$
10$^{-4}$).
With a more recent estimate of the interstellar C/H value (2.6 $\times$
10$^{-4}$, 
Cardelli et al. 1996),  C$_{60}^+$ would contain 0.85 -- 1.7\% of carbon.
This is significant in comparison with the estimated abundance of
PAHs in the interstellar medium, which is $\sim$10 -- 15\% of
cosmic carbon (Joblin et al. 1992, Sellgren et al. 1995, Jones et al. 1996,
Dwek et al. 1997).

More recent observations of the 9577 and 9632 \AA\ DIBs 
(Jenniskens et al. 1997) lead to a lower value of the measured
equivalent width, and thus to a lower abundance range for C$_{60}^+$: 
0.25 to 0.52\% of interstellar carbon (with C/H from Cardelli et al. 1996). 
The reason for this disagreement is 
a different evaluation of the telluric contribution in the DIB profiles.
Maier (1994) and Jenniskens et al. (1997) are both very cautious concerning 
the  C$_{60}^+$ identification,
on the basis of: i) the absence of weaker bands in the ISM near 9400 \AA, 
which are expected from the laboratory spectrum;
ii) the difference in the intensity ratio of the two bands in 
the laboratory and in interstellar spectra;
and iii) the larger shift in wavelength from the Neon matrix to the ISM
compared to the wavelength shift between
rare-gas matrices and the gas phase measured in similar molecules.

As an attempt to clarify this situation, 
we propose an alternate determination of the C$_{60}^+$ abundance 
through its vibrational emission bands. 
Ionization models (Bakes and Tielens 1995) predict
the neutral C$_{60}$ abundance 
to be of the order of 0.7 times the C$_{60}^+$ abundance. 
 
\begin{figure}
\epsfxsize=14cm
\caption{The SWS06 spectrum of the reflection nebula NGC 7023 
in the wavelength range 7.0 -- 8.7 $\mu$m at the maximum resolution 
of $\lambda / \Delta \lambda$ = 1900.
The locations of the C$_{60}^+$ emission bands measured in the laboratory
(Fulara et al.  1993) are marked by vertical lines. 
The spectrum is the mean value of upwards and downwards scans, 
rebinned at a resolution of 0.0042 $\mu$m.
One-half the difference between upwards and downwards
scans reduced independently is also plotted to provide an estimate of the noise.
The minor mismatch at around 7.9 $\mu$m is due to the split of the 
spectrum into two subscans (see section 4).
The calibration curve of one typical detector (number 18) is also plotted.
At 8.02 $\mu$m, we detect the 0--0 S(4) rotational line of molecular
hydrogen, which will be addressed in a future paper (Sellgren et al. 1999).}
\centerline{\hbox{\epsffile{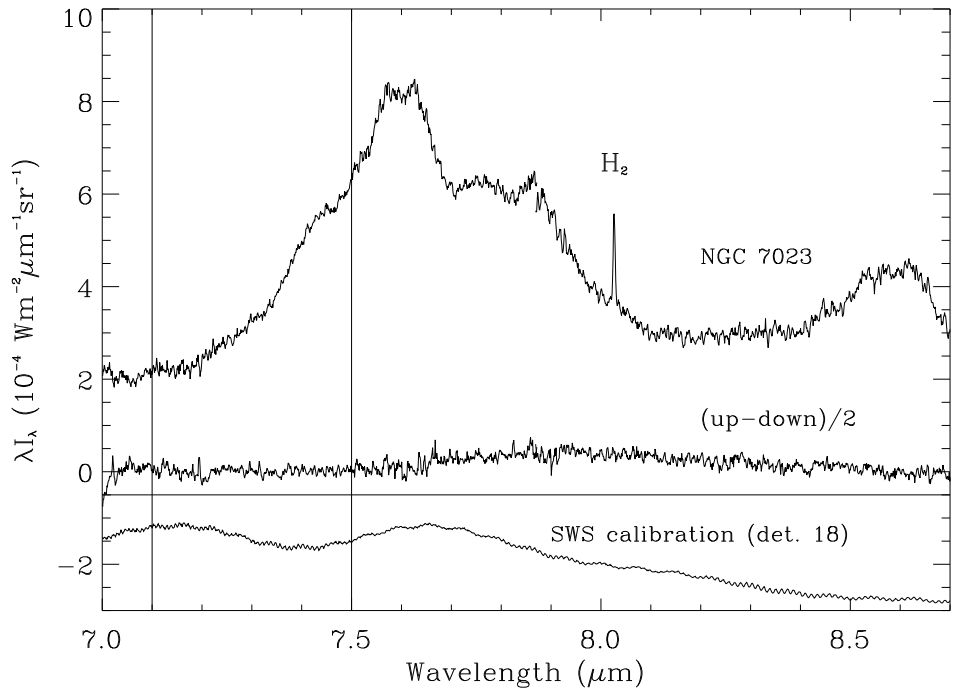}}}
\end{figure}

\section{C$_{60}^+$ in the infrared}

The IR laboratory spectrum of C$_{60}^+$ has been measured in a
Neon matrix (Fulara et al. 1993) and exhibits two bands at 7.11 and 7.51 $\mu$m 
(1406 and 1331 cm$^{-1}$).
We reproduce the laboratory spectrum in Figure 2, courtesy of the authors.
Two other bands are expected for symmetry reasons, 
but they have not been measured so
far because their wavelengths are longer than the 
16.7 $\mu$m (600 cm$^{-1}$) cutoff of the Fulara et al. experiment.
The C$_{60}^+$ features have widths of 0.05 $\mu$m
(10 cm$^{-1}$) in the Neon matrix at low temperature.
The ratio of the integrated cross-section of the
7.11 $\mu$m band to the integrated cross-section
of the 7.51 $\mu$m band is measured to be 2.2 (Fulara et al. 1993).

The modes of neutral C$_{60}$ are already known to lie at 
7.0 and 8.45 $\mu$m (1435 and 1190 cm$^{-1}$; Kr\"atschmer et al. 1990). 
Temperature-dependence 
studies have shown that at 1000 K a shift of $\sim$20 cm$^{-1}$ 
could be expected 
(Nemes et al. 1994), compared to the cold neon matrix.

\begin{figure}
\epsfxsize=11cm
\caption{The vibrational laboratory spectrum of 
C$_{60}$ neutral and ionized species 
in the 6 -- 8 $\mu$m wavelength range, from a Neon matrix experiment 
(by courtesy of Fulara et al. 1993). }
\centerline{\hbox{\epsffile{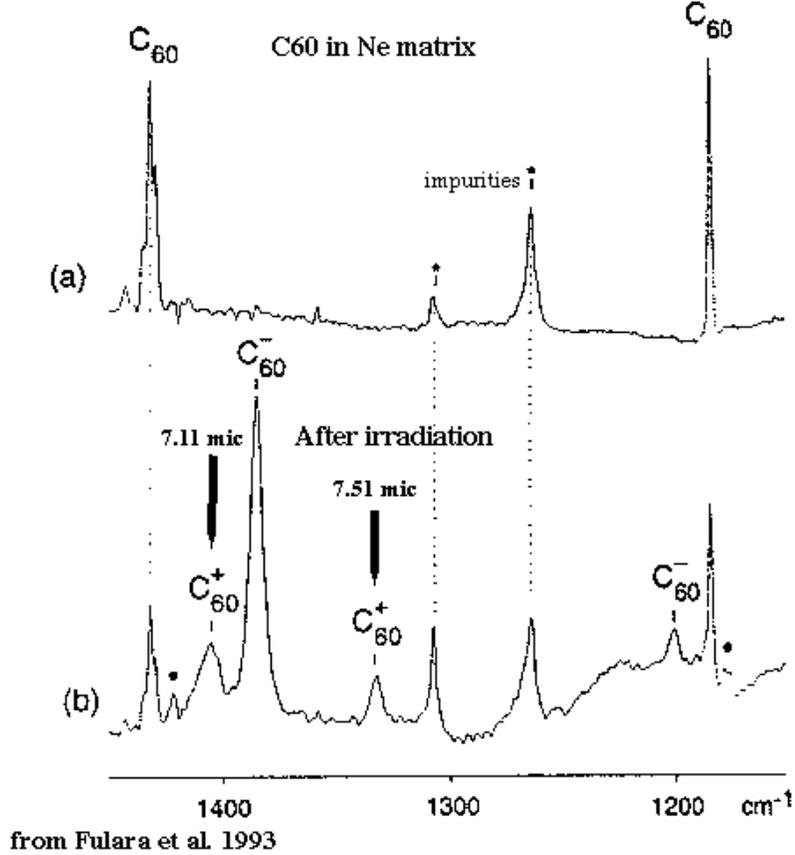}}}
\end{figure}

In a similar way, Joblin et al. (1995) have shown that the widths of PAH 
features measured in emission at high temperature in the laboratory
are broadened by 0 to 0.1 $\mu$m (0 to 20 cm$^{-1}$) 
compared to absorption bands in a 4 K Neon matrix.
They also emphasize that the observed wavelengths of PAH features shift
to longer wavelength with increasing emission temperature. 
The amount of the wavelength shift varies between different vibrational modes
of a PAH, but the redshift is in the range of
0 to 0.1 $\mu$m (0 to 20 cm$^{-1}$) compared to PAHs at
low temperature. No experimental studies have been
published for ionic PAH species. 
It can nevertheless
be expected that C$_{60}^+$ features are broadened and 
shifted by 0 to 20 cm$^{-1}$ as well.
The two C$_{60}^+$ features we wish to detect are therefore expected to lie at 
7.1 -- 7.2 $\mu$m and at 7.5 -- 7.6 $\mu$m. 

We have considered here two cases which are illustrated in Figure 3. 
In Case A, we assume that other carbonaceous species, such 
as PAHs or very small grains, 
emit the 6.2, 7.7, and 8.6 $\mu$m emission features,
and that the combined emission of these features
can be treated as a local continuum.
We will not address here the nature of these emitters. 
For Case A, we have searched for C$_{60}^+$ features at 
7.1 -- 7.2 $\mu$m and at 7.5 -- 7.6 $\mu$m
as weak features above the local continuum
defined by the blue side of the broad 7.7 $\mu$m feature profile.
In Case B, we assume that
the broad 6.2, 7.7, and 8.6 $\mu$m features are each made of 
many overlapping bands from individual aromatic molecules, and 
that two of the bands contributing to this emission
complex could be due to C$_{60}^+$.
The deduced abundance for C$_{60}^+$ would be 
dramatically different between these two cases,
as is shown in Figure 3. 
For both Case A and Case B, 
we adopt Gaussian profiles with a FWHM of 0.05 $\mu$m
for the two C$_{60}^+$ bands, 
and we place
the constraint on any possible C$_{60}^+$ features that the 
ratio of the integrated cross-section of the
7.11 $\mu$m band to the integrated cross-section
of the 7.51 $\mu$m band equal the laboratory
value of 2.2 (Fulara et al. 1993).
This constraint limits the maximum contribution of C$_{60}^+$, 
by requiring that in Case B the stronger of the two C$_{60}^+$
features not exceed the total observed flux at 7.11 $\mu$m
in NGC 7023.

\begin{figure}
\epsfxsize=12cm
\caption{This plot shows the two cases considered here for 
extracting an upper limit on the
C$_{60}^+$ abundance in NGC 7023.
In Case A (top), C$_{60}^+$ features at 7.11 and 7.51 $\mu$m
appear above a local continuum defined by the blue
side of the 7.7 $\mu$m feature profile. 
The best fits to the data, consisting of
second-order polynomial continua
plus Gaussians, are shown. 
In Case B (bottom),
C$_{60}^+$ features at 7.11 and 7.51 $\mu$m
are two of many individual aromatic bands blended
together to produce emission at 7.0 -- 7.8 $\mu$m.
For both cases, we require that the
ratio of the integrated cross-section of the
7.11 $\mu$m band to the integrated cross-section
of the 7.51 $\mu$m band
equal the laboratory value of 2.2 
and the FWHM of each feature be 0.05 $\mu$m
(Fulara et al. 1993).
}
\centerline{\hbox{\epsffile{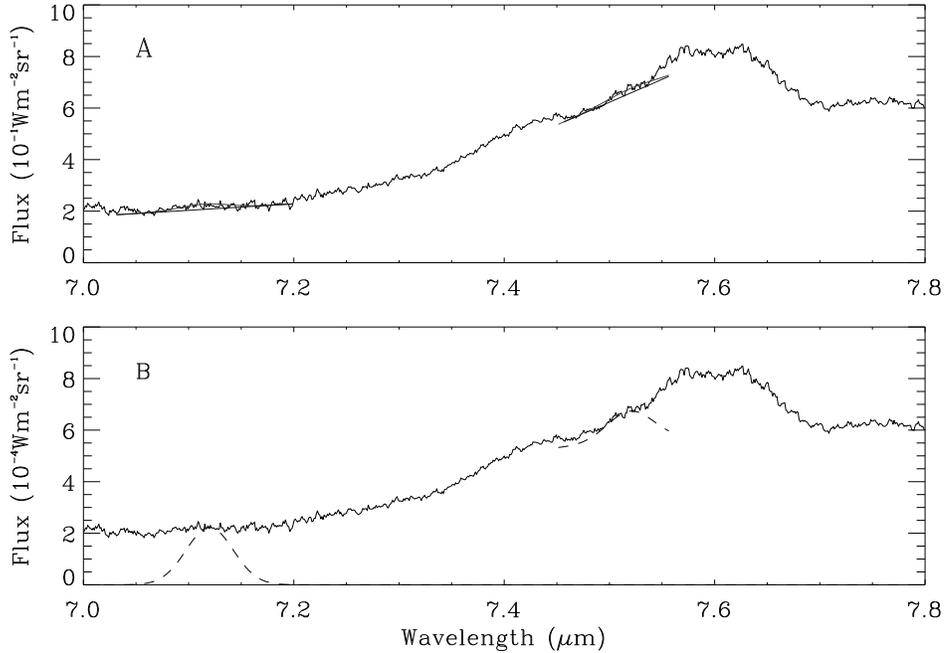}}}
\end{figure}

\section{Observations and data reduction}

The exciting star of the reflection nebula NGC 7023 is the Be star HD 200775, 
which has an effective temperature of 17000 K (Strom et al. 1972). 
For detailed reviews of previous studies of this object, see 
Lemaire et al. (1996) and Martini et al. (1997). 
Our SWS observations are higher spectral resolution than previously
published mid-IR spectra of NGC 7023 
from the Kuiper Airborne Observatory (Sellgren et al. 1985)
and from ISO with ISOPHOT-S and ISOCAM
(Laureijs et al. 1996; Cesarsky et al. 1996). 

The data presented here were taken by the SWS spectrometer on board the 
ISO satellite (Kessler et al., 1996) during revolution 455 (February 13th,
1997). We used the partial grating scan mode (AOT6, band 2C) 
of the SWS instrument (de Graauw et al. 1996). 
The position angle of the major axis of the SWS 
rectangular aperture (14$''$ $\times$ 20$''$ for the wavelength range 
observed here) was 258.5$^\circ$. 
The aperture is thus located at the southern 
edge of the northwestern filament observed by 
ISOCAM (as seen in the 6.2 to 11.3 $\mu$m feature ratio, Cesarsky et al., 1996)
and is parallel in orientation to this filament.
In the SWS grating spectrometer, light is dispersed along the major axis. 

Data gathered by the SWS spectrometer have a high redundancy: 12 detectors 
and two grating scan directions (UP and DOWN hereafter),
for a total of 24 individual spectral scans.
There are two grating steps per spectral resolution element 
($\Delta \lambda = 0.0042 \ \mu$m); each spectral resolution element
is thus observed independently 48 times. 

To reduce the data, we used the current SWS-IA version running at the Institut
d'Astrophysique Spatiale (last update: October 1997). The data reduction steps
are summarized in de Graauw et al. (1996). The flux calibration was done with 
the file CAL25-2C-020. In the following, we detail our modifications to the 
standard reduction scheme in order to improve the signal-to-noise (S/N) 
ratio.

First, we removed glitches due to cosmic rays from the spectra. 
For the band 2C detectors, glitches rarely affect more than a single read-out 
(2 seconds). To detect glitches, we subtract a
continuum (a sliding boxcar on 15 points, with 2-$\sigma$ outliers
rejected) from each of the 24 individual scans, 
to create individual high-frequency (HF) scans. 
The individual HF scans, 
which now contain unresolved lines, glitches and noise,
are combined by wavelength. 
The combined HF spectrum has 48 points per resolution element;
we smooth it with a 15-point sliding boxcar to improve our S/N on the
combined HF signal while retaining enough spectral sampling to detect
true HF features in the spectrum such as unresolved emission lines.
We subtract the combined HF signal from each HF individual 
scan, leaving individual residual scans
which in principle only contain noise and glitches. We finally remove
glitches from each individual residual scan by 3-$\sigma$ clipping, with
$\sigma$ determined locally from a 10-point sliding boxcar.
Between 10 and 20\% of the data points were identified as glitches.

Second, we removed correlated noise, due to weaker glitches 
(not necessarily due to cosmic rays like, e.g., electrical noise)
which occur in all 12 detectors at the same time.
Correlated noise is not always detected by 
the above glitch removal technique, and
can result in a false spectral feature which appears broader than 
the spectral resolution (see Sellgren et al. 1999).
To eliminate correlated noise, we subtracted a continuum from each scan
to create individual HF scans, as described above, but then
combined the HF scans in the time domain (at each reset point)
rather than in the wavelength domain. 
We next subtracted from each individual HF scan
the mean HF signal (after sigma-rejection) of 
all 12 detectors taken at each reset point. The 
individual continuum is subsequently restored into each corrected
HF scan. Unresolved lines remain
unaffected by this method as our sampling is dense enough. By applying 
this signal decorrelation, we reduced the rms noise by 5 to 15\%.

The two operations described above, to correct for glitches
and correlated noise, were performed just after flux calibration. 
As they were performed on the HF part of the signal they do not depend 
on flatfield quality. 

The third step in the data reduction is the flatfielding. 
We used the standard SWS flatfielding routine to scale all 
individual scans to a reference. The reference is defined by a 
linear fit to all 12 DOWN scans. 
The scaling factor for each individual scan is defined
as the reference divided by a linear fit to the individual scan.
If an individual scan has some fit points close to zero, 
the scaling factor becomes large and  
the amplification of noise in the scan adversely affects
the flatfield quality. To avoid this, we added a constant value (100 Jy)
to all individual scans, performed the flatfield correction
(multiplication by the scaling factors), then subtracted this
constant value to produce the final flux level. 
The final average spectrum was produced with a sliding boxcar on a given 
binsize (0.0042 $\mu$m or mean resolving power of 1900, Valentijn et al., 
1996).

In order to be sure that the
residual error includes any true differences between the UP and DOWN scans, 
we performed a second reduction in which we separated the UP and DOWN 
scans just before the flatfield operation. 
The two scan directions were then flatfielded completely 
independently. This was used only to calculate 
the residual error (UP $-$ DOWN)/2. 

\section{Estimate of the C$_{60}^+$ upper limit}

We will now derive an upper limit 
on the C$_{60}^+$ and C$_{60}$ abundances, for 
both Case A and Case B.
For this purpose, we compare the observed spectrum of NGC 7023 to a 
calculated emission spectrum of C$_{60}^+$. 
This is more rigorous than extracting a 
relative value of C$_{60}^+$ emission compared to the PAH emission,
especially in the 7 -- 8 $\mu$m domain, where the 
spectroscopic match of PAHs to the interstellar
data is not yet proved.
We will first show the detailed calculation for C$_{60}^+$, and then apply 
it to C$_{60}$.

\subsection{Assumptions}

The C$_{60}^+$ absorption spectrum we adopt for modeling purposes
reproduces the bandwidths and relative
intensities measured in laboratory experiments (Fulara et al. 1993).
We use the wavelengths and relative intensities of the
long-wavelength modes in C$_{60}$ 
(at 17.3 and 18.9 $\mu$m) to approximate these
modes in C$_{60}^+$, since the long-wavelength modes of C$_{60}^+$ are
predicted but not yet measured.
We assume the wavelengths of the C$_{60}^+$ modes do not depend 
on temperature, because we cannot predict this dependence 
(although any wavelength shift is probably less than 
0.1 $\mu$m). It is important to
note that the power emitted in the C$_{60}^+$ features
does not depend on the absolute value of the IR
cross-section. The reason for this is energy conservation; 
the total absorbed power is equal to the total emitted power,
independent of the cooling time between UV photons.

According to the thermal model (Sellgren 1984, 
L\'eger et al. 1989), the excitation of a species after a single 
UV or visible photon absorption raises its vibrational temperature to a 
high value, determined by its size via the heat capacity.
The emission results from the cooling by vibrational decay. To calculate the IR
emission spectrum of C$_{60}^+$, we thus have to model its absorption of
stellar light and its emission in the 7 $\mu$m bands.

The total power absorbed by  C$_{60}^+$ molecules near a star is given by:
\begin{equation}
 P_{C_{60}^+} = N_{C_{60}^+} \int \ F_{\star} (\nu)  \ 
\sigma_{UV C_{60}^+} (\nu) \ d\nu
\end{equation}
where $F_{\star} (\nu)$ is the illuminating stellar flux
as a function of frequency $\nu$, $\sigma_{UV C_{60}^+}
(\nu)$ is the UV absorption cross-section 
per carbon atom for a C$_{60}^+$ molecule, 
and $N_{C_{60}^+}$ is the number of carbon atoms in the
observed source which are contained within C$_{60}^+$ molecules.

The UV absorption cross-section of the 
C$_{60}^+$ cation is not directly known from laboratory
experiments.  We therefore use the value measured from experiments on
neutral PAHs, normalized to one carbon atom, 
as it seems to be similar for graphite and PAHs and thus
characteristic of sp$^2$ hybridized carbon (Joblin et al. 1992,
Verstraete \& L\'eger 1992). We note that 
the absorption
cross-section for an ionized species is expected to be larger than for a
neutral one, especially in the visible (Leach 1987). Our assumption thus 
probably underestimates
the actual absorption, which is suitable for determining an upper limit of the
species abundance. 

The emitted spectral energy distribution, 
$P_{em} (\nu)$, 
depends on the infrared absorption cross-section as a function
of wavelength, and the details of how the molecule cools after
absorption of a visible or UV photon,
as described by L\'eger et al. (1989).
We define
$f_{7 \mu {\rm m}}$, the fraction of the total C$_{60^+}$ 
emitted energy which is emitted 
in the C$_{60^+}$ bands at 7.11 and 7.51 $\mu$m, as
\begin{equation}
f_{7 \mu {\rm m}} \ = \ \frac{\int_{\nu_1}^{\nu_2} \ P_{em}(\nu) \ d \nu }
{\int_{0}^{\infty} \ P_{em}(\nu)\ d\nu}
\end{equation}
where the frequency limits $\nu_1$ and $\nu_2$
indicate an integration just over the 7.11 and 7.51 $\mu$m bands.
The total C$_{60^+}$ emitted energy will include 
emission in the C$_{60^+}$ bands at 7.11 and 7.51 $\mu$m, 
emission in the C$_{60^+}$ long-wavelength modes 
(approximated by the C$_{60}$ modes at 17.3 and 18.9 $\mu$m), 
and possibly also some unknown amount of fluorescence.
Such fluorescence is observed in some neutral 
aromatic species (Negri \& Zgierski 1994).
The C$_{60}^+$ fluorescence
quantum yield is not known, and so we assume it is negligible.
Including a finite amount of C$_{60}^+$ fluorescence would have
the effect of increasing our upper limit on the
C$_{60}^+$ abundance.
Our calculations, ignoring fluorescence,
find that the fraction of the total emitted
energy from C$_{60}^+$ is  35\% at 7.11 $\mu$m, 15\% at 7.51 
$\mu$m, 10\% at 17.3 $\mu$m, and 40\% at 18.9 $\mu$m. 
This gives $f_{7 \mu {\rm m}}$ =  50\% as the fraction of total C$_{60}^+$ 
emitted energy resulting from emission at 7.11 and 7.51 $\mu$m.

Energy conservation requires that $\int_{0}^{\infty} \ P_{em} (\nu) \ d\nu$ 
= $P_{C_{60}^+}$. 
To convert the emitted power spectrum from C$_{60}^+$ to an observed 
total intensity spectrum, $I_{C_{60^+}} (\nu)$,
we compare the power absorbed and re-emitted in the C$_{60}^+$ bands 
to the power absorbed and re-emitted by dust grains, $P_{dust}$ and $I_{dust}$.
We derive $I_{dust}$ from the integration of the SWS01 and LWS01 spectra of 
NGC 7023 (Sellgren et al. 1999), over the wavelength range 3 -- 200 $\mu$m. 
Our value of I$_{dust}$ given in Table 1,
1.0 $\times$ 10$^{-3}$ W m$^{-2}$ sr$^{-1}$,
is compatible with previous
measurements (Whitcomb et al. 1981).

The power thermally emitted by the entire grain population is related to the 
total stellar flux absorbed by dust by:
\begin{equation}
P_{dust} = N_H \int (1-\omega(\nu)) 
\ F_{\star}(\nu) \ \sigma_{UV dust}(\nu)\  d\nu 
\end{equation}
where $N_H$ is the total number of 
hydrogen atoms, 
$\sigma_{UV dust}$ is the grain extinction cross-section
normalized to one hydrogen atom,
and $\omega$ is the grain albedo.
We adopt the extinction curve of the 
diffuse interstellar medium and the associated
albedo as given in D\'esert et al. (1990). 
We integrated over ten points in the UV-visible range.

The fraction of carbon locked up in 
C$_{60}^+$, $f_{C_{60^+}}$, is then given by:
\begin{equation}
f_{C_{60^+}} = \left ( {f_{7 \mu \rm m}}  \right ) ^{-1} 
\ \frac{P_{dust}\ /\ N_H}{P_{C_{60}^+}\ /\ N_{C_{60}^+}} \ 
\left ( \frac{N_C}{N_H} \right ) ^{-1} \ \frac{I_{C_{60}^+}}{I_{dust}} 
\end{equation}
The first term in this equation 
is $f_{7 \mu \rm m}$, the fraction of total C$_{60}^+$ energy 
emitted at 7.11 and 7.51 $\mu$m.
The second term is the ratio of stellar flux absorbed
per hydrogen atom by dust
to the stellar flux absorbed 
per carbon atom by C$_{60}^+$.
This term is effectively the ratio of 
the dust absorption cross-section 
to the C$_{60}^+$ absorption cross-section,
weighted by the stellar flux.
The stellar flux is 
approximated by a black body emission spectrum
at the effective temperature of HD 200775, i.e. 17,000 K.
The third term is the interstellar carbon abundance,
$N_{C}/N_H$, 
as measured by Cardelli et al. (1996) 
in the local interstellar medium.
The last term is our observed upper limit on the 7.11 and 7.51 $\mu$m
C$_{60}^+$ emission, $I_{C_{60}^+}$, divided by the observed total
infrared emission from dust, $I_{dust}$.
Our adopted values for all these terms
used in Equation 4
are summarized in Table 1.

\begin{table}
\caption{Summary of calculated quantities and observables which lead to 
an upper limit on the C$_{60}^+$ abundance in NGC 7023.}
\begin{center}
\begin{tabular}{ll}
\hline
$P_{C_{60}^+}/N_{C_{60}^+}$ & 7.7 $\times$ 10$^{-24}$ W C$^{-1}$\\
$P_{dust}/N_H $& 1.35 $\times$ 10$^{-27}$ W H$^{-1}$\\
$I_{dust}$ & 1.0 $\times$ 10$^{-3}$ W m$^{-2} \ {\rm sr}^{-1}$ \\
$I_{C_{60}^+}$ (Case A) & 
$<$2.0 $\times$ 10$^{-7}$ W m$^{-2} \ {\rm sr}^{-1}$ \\
$I_{C_{60}^+}$ (Case B) & 
$<$1.9 $\times$ 10$^{-6}$ W m$^{-2} \ {\rm sr}^{-1}$ \\
$N_{C}/N_H$ & 2.6 $\times$ 10$^{-4}$ \\
\hline
\end{tabular}
\end{center}
\end{table}

\subsection{Case A}
Let us consider Case A, as presented in the top panel
of Figure 3, where the two C$_{60}^+$ bands
are small bumps on top of a local continuum.
We adopt a second-order local continuum and we fit each small bump with
a Gaussian of 0.05 $\mu$m FWHM.
The residual emission at 7.11 and 7.51 $\mu$m, after continuum subtraction,
is only significant at the
2.5-$\sigma$ level and thus we consider this
to be an upper limit on the
maximum integrated energy 
of any C$_{60}^+$ features. 
We estimate this upper limit to be 
$<$2.0 $\times$ 10$^{-7}$ W m$^{-2}$ sr$^{-1}$ for the sum of both bands.
The deduced upper limit for Case A is $<$0.027\% of carbon atoms contained in
C$_{60}^+$ towards NGC 7023 (equation 4). 

\subsection{Case B}
We now consider the other extreme, Case B, where 
we place an upper limit on C$_{60}^+$ 
as illustrated in Figure 3b.
We again adopt Gaussians of width similar to those measured in the 
laboratory (FWHM $=$ 0.05 $\mu$m), 
for both the 7.11 and 7.51 $\mu$m bands,
and require that the 
ratio of the integrated cross-section of the
7.11 $\mu$m band to the integrated cross-section
of the 7.51 $\mu$m band equal the laboratory
value of 2.2 (Fulara et al. 1993).
These requirements for our Case B upper limit imply that
all the emission at 7.11 $\mu$m, but only part of the
emission at 7.51 $\mu$m, is due to C$_{60}^+$
in the observed spectrum of NGC 7023.

For Case B, the upper limit on the total energy 
emitted by C$_{60}^+$ at 7.11 and 7.51 $\mu$m 
is $<$1.9 $\times$ 10$^{-6}$ W m$^{-2}$ sr$^{-1}$.
This leads to an upper limit on the 
C$_{60}^+$ abundance of $<$0.26\% of the interstellar carbon
in NGC 7023. 

\subsection{Comparison of C$_{60}^+$ abundances from different methods}

Our Case B method gives an upper limit on the C$_{60}^+$ abundance
($<$0.26\% of interstellar carbon)
which is just compatible with the C$_{60}^+$ abundance 
(0.25 -- 0.52\% of interstellar carbon)
inferred from the DIBs observed at 9577 and 9632 \AA\ 
by Jenniskens et al. (1997).
Our Case B results, however, are
significantly lower than the C$_{60}^+$ abundance
(0.85 -- 1.7\% of interstellar carbon) inferred from the
DIB observations of Foing and Ehrenfreund (1994, 1997).
Our upper limit on the C$_{60}^+$ abundance from Case A 
($<$0.027\% of interstellar carbon) 
is a factor of ten below
the C$_{60}^+$ abundances inferred from any of the
DIB observations.
We note that the identification of the
9577 and 9632 \AA\ DIBs with C$_{60}^+$, however,
is still debated (Foing and Ehrenfreund 1994, 1997;
Maier 1994; Jenniskens et al. 1997).
We also note that if the C$_{60}^+$ fluorescence
quantum yield, $Q_f$, is not negligible, it would raise both our
Case A and Case B upper limits by a factor of
(1 $-$ $Q_f$)$^{-1}$.

In Case B, we assume that
C$_{60}^+$ is one of many narrow (FWHM $\sim$ 0.05 $\mu$m)
aromatic bands blended
together to emit features at 6 -- 9 $\mu$m,
as has been proposed in the PAH model
(L\'eger \& Puget 1984, Allamandola et al. 1985). 
If this view of the origin of the
interstellar emission features at 6 -- 9 $\mu$m
is correct, then this has the interesting implication 
that $\sim 10^2$ different aromatic species
would be required to account for the entire 6 -- 9 $\mu$m
emission complex.
It is not clear, however, whether such a mix of aromatic
molecules could also account for the
underlying smooth continuum at 1 -- 14 $\mu$m,
particularly in wavelength regimes where PAHs do not have
vibrational transitions.

Our Case B upper limit on the C$_{60}^+$ abundance is
set simply by the total observed flux at 
7.11 $\mu$m in NGC 7023.
If there are additional aromatic molecules which contribute
to the observed flux at 7.11 $\mu$m, as seems likely,
then our upper limit to the C$_{60}^+$ abundance of
$<$0.26\% of interstellar carbon is an overestimate.

\subsection{Neutral C$_{60}$}
We also searched for the neutral C$_{60}$ features near 7.0 -- 7.1
and 8.45 -- 8.6 $\mu$m (Kr\"atschmer et al. 1990, 
Nemes et al. 1994). 
We found no resolved emission features at the expected wavelengths. 
We estimate an upper limit on the neutral C$_{60}$ abundance
using the Case B method,
by assuming that all of the the observed flux in NGC 7023
at 7.0 $\mu$m is due to C$_{60}$.
We require that the ratio of the integrated cross-section of the
8.45 $\mu$m band to the integrated cross-section
of the 7.0 $\mu$m band matches 
the experimental ratio of 0.75, 
and we assumed that the features were Gaussians
with a width of 0.05 $\mu$m.
We find an upper limit on 
$I_{C_{60}}$ (Case B) of $<$2.0 $\times$ 
10$^{-6}$ W m$^{-2} \ {\rm sr}^{-1}$.
We follow the same procedure for C$_{60}$ as for
C$_{60}^+$, as our calculation did not make any
corrections for the ionization state of the fullerene.
This implies a Case B upper limit for the abundance of the
neutral C$_{60}$ molecule of
of $<$0.27\% of interstellar carbon.

Ionization models predict that
the neutral C$_{60}$ abundance
should be $\sim$0.7 times the C$_{60}^+$ abundance
in the highly illuminated environment of a B star (Bakes and Tielens 1995). 
Our Case B upper limits on the abundances of
C$_{60}$ and C$_{60}^+$,
respectively $<$0.27\% 
and $<$0.26\% of interstellar carbon,
do not contradict the predictions of ionization models,
as we do not detect either molecule in NGC 7023.

\section{Conclusions}
We have searched for the vibrational emission bands 
of C$_{60}^+$, at 7.11 and 7.51 $\mu$m, 
and of C$_{60}$, at 7.0 and 8.45 $\mu$m, 
using a new ISO spectrum of the reflection nebula NGC 7023.
We do not detect emission from either fullerene molecule,
and we place upper limits on their abundances.
Our upper limit on the C$_{60}^+$ abundance,
from the lack of observed vibrational emission,
is independent of other estimates based on
assigning diffuse interstellar bands at 9577 and 9632 \AA\ 
to electronic absorption bands of C$_{60}^+$
(Jenniskens et al. 1997, Foing \& Ehrenfreund 1994, 1997). 

If we assume that all of the flux observed toward
NGC 7023 at 7.11 $\mu$m is due to C$_{60}^+$, then
we derive an upper limit on the
C$_{60}^+$ abundance of $<$0.26\% of interstellar
carbon toward NGC 7023.
This upper limit is just
compatible with the Jenniskens et al. (1997)
observations, which imply 0.25 to 0.52\%
of interstellar carbon is in C$_{60}^+$,
and is significantly below the C$_{60}^+$ abundance
inferred from Foing \& Ehrenfreund (1994, 1997).
If we instead assume that C$_{60}^+$ would produce
only a small bump on a local continuum due to the
blue wing of the 7.7 $\mu$m interstellar emission
feature, then our upper limit on the C$_{60}^+$
abundance, $<$0.027\% of interstellar carbon,
is a factor of ten less than the estimates
from the 9577 and 9632 \AA\ DIBs.

The apparent lack of C$_{60}^+$ toward NGC 7023
is unlikely to be due to the ionization
state of the fullerene.
Our upper limit on C$_{60}$, $<$0.27\% of
interstellar carbon, is comparable to our upper
limit on C$_{60}^+$.
Ionization models of aromatic hydrocarbons predict
that the fraction of C$_{60}^{++}$ is negligible in NGC 7023 
(Bakes \& Tielens 1995). 
These models also predict that the C$_{60}^+$
fraction should be higher in NGC 7023 
than in the diffuse medium where
Foing \& Ehrenfreund (1994, 1997) and Jenniskens et al. (1997)
made their observations.

There are several possible explanations for why
C$_{60}^+$ is not detected toward NGC 7023 in the
amount inferred from the near-IR DIB observations.
It is possible that we have overestimated the fraction
of energy absorbed by C$_{60}^+$ which is re-radiated
near 7 $\mu$m, for instance if the C$_{60}^+$ far-infrared
modes are stronger than expected from data on C$_{60}$,
or if fluorescence is non-negligible.
Also, while the 9577 and 9632 \AA\ DIBs have not yet
been observed toward the central star in NGC 7023,
other DIBs are observed to be anomalously weak toward this star 
(Snow et al. 1995; Oudmaijer et al. 1997).
Jenniskens et al. (1997) observe that 
the 9577 and 9632 \AA\ DIBs follow the behavior of
other DIBs (Jenniskens et al. 1994) 
in being significantly fainter 
in dense molecular clouds with strong
UV irradiation. 
A third explanation is that the 9577 and 9632 \AA\ DIBs
are not due to C$_{60}^+$, as argued by
Maier (1994) and Jenniskens et al. (1997).

New laboratory data will be critical to answering some of these questions.
It is essential to obtain the near-IR gas-phase spectrum 
(0.94 to 1.0 $\mu$m) of C$_{60}^+$, in order to
confirm or deny the assignment of the 9577/9632 \AA\ DIBs to C$_{60}^+$.
It is important to measure the broadening and wavelength shift
induced by temperature effects on the vibrational bands of C$_{60}^+$,
which we can only estimate from laboratory measurements on similar molecules.
It would also be useful to observe the full-range IR 
absorption spectrum of C$_{60}^+$ to measure the relative cross-sections
of all four vibrational bands.

We conclude that C$_{60}$ related species do not appear 
to be a dominant form of carbonaceous 
species in the interstellar medium. 
Confirmation of this result, as well as further searches for fullerene or 
fullerane species in interstellar spectra, should be pursued at various 
wavelengths. 
Additional searches for the fullerene bands near 7 $\mu$m in
circumstellar environments,
where C$_{60}$ and related molecules could be formed,
could be particularly fruitful. \\[.5cm]

{\bf Acknowledgements}: The Infrared Space Observatory is a 
project of the European Space Agency.
We particularly thank the SWS operational team at Vilspa for their efforts
to provide the calibration and data reduction systems.
We gratefully acknowledge NATO Collaborative Research Grant 951347.

\end{document}